\documentclass[secnumarabic, graphics,floatfix, nofootinbib,tightenlines,nobibnotes, aps, prl, 12pt]{revtex4-1}
\usepackage{graphicx}
\usepackage[english]{babel}
\usepackage{mathrsfs}
\usepackage{scalerel,amsmath,amssymb}
\usepackage{amsfonts}
\usepackage{multirow}
\usepackage[section]{placeins}

\usepackage[center]{subfigure}

\begin{document}

 \newcommand{\bq}{\begin{equation}}
 \newcommand{\eq}{\end{equation}}
 \newcommand{\bqn}{\begin{eqnarray}}
 \newcommand{\eqn}{\end{eqnarray}}
 \newcommand{\nb}{\nonumber}
 \newcommand{\lb}{\label}

\title{On matrix method for black hole quasinormal modes}

\author{Kai Lin$^{1,2}$}\email{lk314159@hotmail.com}
\author{Wei-Liang Qian$^{2,3,4}$}\email{wlqian@usp.br}

\affiliation{$^{1}$ Hubei Subsurface Multi-scale Imaging Key Laboratory, Institute of Geophysics and Geomatics, China University of Geosciences, 430074, Wuhan, Hubei, China}
\affiliation{$^{2}$ Escola de Engenharia de Lorena, Universidade de S\~ao Paulo, 12602-810, Lorena, SP, Brazil}
\affiliation{$^{3}$ Faculdade de Engenharia de Guaratinguet\'a, Universidade Estadual Paulista, 12516-410, Guaratinguet\'a, SP, Brazil}
\affiliation{$^{4}$ School of Physical Science and Technology, Yangzhou University, 225002, Yangzhou, Jiangsu, China}

\date{Feb. 26, 2019}

\begin{abstract}
In this paper, we provide a comprehensive survey of possible applications of the matrix method for black hole quasinormal modes.
The proposed algorithm can generally be applied to various background metrics, and in particular, it accommodates for both analytic and numerical forms of the tortoise coordinates, as well as black hole spacetimes.
Our discussions give a detailed account of different types of black hole metrics, master equations, and the corresponding boundary conditions.
Besides, we argue that the method can readily be applied to cases where the master equation is a system of coupled equations. 
By adjusting the number of interpolation points, the present method provides a desirable degree of precision, in reasonable balance with its efficiency.
The method is flexible and can easily be adopted by various distinctive physical scenarios.
\end{abstract}


\maketitle

\section{I. Introduction}
\renewcommand{\theequation}{1.\arabic{equation}} \setcounter{equation}{0}

The black hole is one of the most exotic and intriguing subjects in all of theoretical physics.
On the experimental side, black hole merger is a magnificent astrophysical phenomenon, it releases an enormous amount of energy in the form of gravitational radiation, even more than the light from all the stars from the entire visible Universe combined.
The gravitational waves traverse the Universe, carrying with it the information on the merger.
The first detection of the gravitational wave in 2016 was heralded as a remarkable breakthrough in fundamental physics~\cite{agr-LIGO-01,agr-LIGO-02}.
The observation was announced by LIGO and Virgo collaborations, and it matches the predictions of gravitational waves emanated from the inward spiral and merger of a pair of black holes~\cite{agr-merger-01,agr-merger-02,agr-merger-03}. 
Subsequently, further measurements were confirmed regarding gravitational waves of compact-object binaries, including black holes and neutron stars~\cite{agr-LIGO-03,agr-LIGO-04,agr-LIGO-05}.
The observations inaugurate a revolutionary era of astronomy, where astrophysicists and cosmologists are now able to make precise observations based upon gravitational wave, besides electromagnetic radiation.
Indeed, while black hole candidates have been identified through electromagnetic signals from nearby matter~\cite{agr-obs-general-relativity-01,agr-obs-general-relativity-02,agr-obs-general-relativity-03,agr-obs-general-relativity-04}, gravitational wave measurement provides a direct test of general relativity, as well as unique access to the properties of the spacetime, in the regime of the strong field.
The final fate of a black hole binary consists of three distinctive stages: inspiral, merger, and ringdown~\cite{agr-merger-review-01}.
In the first stage of the inspiral, the orbit of the black hole binary shrinks gradually owing to the emission of gravitational radiation.
The inspiral dynamics can be calculated by using the post-Newtonian approximation, which results from a systematic expansion of the full Einstein equations.
As the two black holes further evolve and spiral inward, they eventually reach the merger stage, which relates to the strong field, the dynamical regime of general relativity.
During the merger stage, the two black holes coalesce into a single, highly distorted remnant black hole, which is surrounded by the combined event horizon. 
The process is no longer quasi-adiabatic and incredibly sophisticated, where most analytic or semi-analytic methods break down, and numerical relativity simulations are usually employed to calculate the dynamical properties of the system. 
Gravitational wave emission reaches a peak during the second stage.
The remnant black hole eventually settles down into a more dormant state, where the distortions from the spherical shape rapidly reduce while emitting gravitational waves.
This last stage is known as ``ringdown", in analogy to how the ringing of a bell is suppressed in time after it is struck.
The form of the gravitational wave is characterized by its shape of exponentially damped sinusoids.
Various techniques concerning the black hole perturbation theory can be used to tackle the problem.
The topic of the present paper is focused on the black hole quasinormal modes, which is closely related to the last stage of the black hole merger process.

In general, quasinormal mode is defined as an eigenmode of a dissipative system.
In the context of general relativity, small perturbations of a black hole metric may lead to quasinormal modes~\cite{agr-qnm-04,agr-qnm-review-01,agr-qnm-review-02,agr-qnm-review-03}.
Its importance is closely associated with the final fate of the black hole merger.
Besides, quasinormal modes arouse much attention in the recent year mainly owing to the development of the holographic principal regarding the anti-de Sitter/conformal field theory (AdS/CFT) correspondence, which is widely recognized as a tool for exploring the strongly coupled systems~\cite{agr-qnm-holography-review-01}.
In other words, by studying the black hole perturbations in the bulk, one may extract important physical quantities, such as the transport coefficients, such as viscosity, conductivity, and diffusion constants, of the dual system on the boundary.

The study of black hole quasinormal modes encompass various types of perturbations, which includes those of scalar field, spinner field, vector field, or the metric itself.
Manifested in terms of the temporal evolution of the small perturbations, the dynamic process for stable black hole metric also involves three stages.
These include the initial outburst triggered by the source of the perturbation, the quasi-normal oscillations, and late-time asymptotic tail.
Usually, the amplitude of the oscillations decays exponentially in time.
Otherwise, if small oscillations increase, it indicates that the black hole metric is not stable.
The frequency of a quasinormal mode is usually complex and therefore is known as quasinormal frequency.
The real part of the quasinormal frequency indicates the frequency of the oscillations.
The imaginary part, on the other hand, is negative for a stable metric as it plays the role to suppress the amplitude of the oscillations.

From a mathematical viewpoint, the analysis of the quasinormal mode is related to a non-Hermitian eigenvalue problem regarding a system of coupled linear differential equations with the associated boundary conditions.
Besides a few cases where analytic solutions for asymptotic quasinormal mode spectra are obtained~\cite{agr-qnm-02,agr-qnm-05,agr-qnm-06}, one often resorts to semi-analytic techniques~\cite{agr-qnm-review-04}.
The P\"oshl-Theller potential approximation~\cite{agr-qnm-Poshl-Teller-01} is to approximate the effective potential in the master equation to P\"oshl-Theller one where analytic eigenvalue is known. 
The WKB method~\cite{agr-qnm-WKB-01,agr-qnm-WKB-02} is a semi-analytic approach proposed by Schutz {\it et.al} where the technique is applied specifically to the one-dimensional time-independent Schr\"odinger-like equation for the quasi-eigenvalue problem.
The 6th order formalism of the method was subsequently derived by Konoplya~\cite{agr-qnm-WKB-03}.
In practice, the application of the WKB method is rather straightforward, since it only requires the information on the background metric and the form of the effective potential.
However, the uncertainty of the obtained result is somewhat undetermined beforehand, which is in part related to the specific shape of the effective potential.
It is worth mentioning that, in the eikonal limit, the quasinormal modes of spherical black holes in asymptotically flat spacetimes were found to be associated with the parameters of the circular null geodesic~\cite{Cardoso:2008bp}.
However, Konoplya and Stuchlik~\cite{Konoplya:2017wot} showed analytically that the relationship is not valid for some particular cases, which has been further explored~\cite{Konoplya:2017lhs,Toshmatov:2018ell,Toshmatov:2018tyo}.
In addition to the semi-analytic method, various numerical methods also have been proposed.
The first type of numerical method aims to evolve the time-dependent wave equation directly for given sets of the initial condition.
It is achieved by replacing the derivative by the finite difference~\cite{agr-qnm-finite-difference-01,agr-qnm-finite-difference-02,agr-qnm-finite-difference-03,agr-qnm-finite-difference-04,agr-qnm-finite-difference-05,agr-qnm-finite-difference-06}.
Naturally, the resulting waveform contains the information on all the three stages of the temporal evolution of the small perturbations.
At the end of the calculations, one may further evaluate the quasinormal frequency by using Fourier analysis.
A major disadvantage of the approach is that one cannot obtain the complete spectrum of the quasinormal modes, owing to the fact that usually only a few long-lived modes can be clearly identified.
The second class of numerical schemes involves to Fourier decompose of the perturbed Einstein equations while adopting the appropriate boundary conditions.
The latter assumes that the resultant wave function corresponds to incoming waves on the horizon and outgoing at infinity.
A series expansion is usually carried out for the wave function, and the method is typically iterative, namely, higher order result is derived from and improved upon the lower order ones.
Subsequently, the precision of the method is controlled by the order of the expansion. 
These methods include the continued fraction method~\cite{agr-qnm-continued-fraction-01,agr-qnm-continued-fraction-02,agr-qnm-review-05}, the Horowitz and Hubeny's (HH) method for AdS black hole~\cite{agr-qnm-HH-01}, and asymptotic iteration method~\cite{agr-qnm-asympototic-iteration-01,agr-qnm-asympototic-iteration-02,agr-qnm-asympototic-iteration-03,agr-qnm-asympototic-iteration-04}.
However, for the continued fraction method it becomes less straightforward to derive the desired iterative relation for specific background metrics, the HH method only applies to the asymptotically AdS spacetimes.
Recently, we proposed a matrix method~\cite{agr-qnm-lq-matrix-01,agr-qnm-lq-matrix-02,agr-qnm-lq-matrix-03} where the spatial coordinate is discretized so that the differential equation, as well as its boundary conditions, are transformed into a homogeneous matrix equation.
A vital feature of the method is that the eigenfunction is expanded in the vicinity of all grid points, and therefore the precision of the algorithm is potentially improved.
The method has been employed to study the quasinormal modes of the Schwartzchild black holes in asymptotically flat, (anti-)de Sitter (AdS/dS) spacetimes, as well as Kerr and Kerr-Sen black hole metrics~\cite{agr-qnm-lq-matrix-02,agr-qnm-lq-matrix-03}.
In this survey, we further explore possible applications of the method to more general scenarios.
We provide a detailed account of the procedure on how the algorithm is implemented.
We argue that the implementation of the proposed method does not rely on the analytic form of the tortoise coordinate.
Moreover, the method can be readily applied even to the cases of numerical black hole metric.
Our discussions enumerate black hole metric of different asymptotic behavior, and in particular, where an analytic form of the tortoise coordinate is not at disposal.
Advantages, as well as improvements upon alternative approaches, are also addressed.

The manuscript is organized as follows.
In the following section, we discuss the core algorithm of the matrix method.
In section III, we explore the applications of the method in various scenarios, inclusively for different background metrics and the corresponding boundary conditions.
In particular, we investigate the method of separation of variables for the master equation in asymptotically flat, dS, and AdS spacetimes, while considering different cases such as static, rotational, and extreme black holes.
Moreover, the cases where the master equations is a system of coupled ordinary differential equation are addressed.
In addition, we provide a comprehensive discussion on the code implementation using {\it Mathematica}.
Further discussions and concluding remarks are given in the last section of the paper.

\section{II. The matrix method}
\renewcommand{\theequation}{2.\arabic{equation}} \setcounter{equation}{0}

The primary feature of the matrix method~\cite{agr-qnm-lq-matrix-01,agr-qnm-lq-matrix-02,agr-qnm-lq-matrix-03} which makes it distinct from others lies in the fact that the series expansion of the wavefunction is carried out in the vicinity of a series of points, where the latter do not necessarily distribute evenly in the domain of the wavefunction.
This feature provides flexibility to achieve a reasonable degree of precision while not sacrificing the efficiency.
Roughly speaking, in terms of the above grid point expansion, one discretizes the master equation of the perturbation field, and subsequently, rewrites it together with the boundary conditions in the form of the matrix equation.
The latter can then be handled readily by various algorithms for solving a system of linear equations.

Let us start by discussing the general strategy to apply the method to the standard scenario for the quasinormal modes, where the master equation is of Schr\"odinger-type as follows:
\bqn
\lb{qnmeq}
\frac{d^2}{dr_*^2}\Phi(r)+\left[\omega^2-V(r)\right]\Phi(r)=0 ~,
\eqn
where $r_*=\int{dr/f(r)}$ is the tortoise coordinate, $\omega$ and $V(r)$ are the quasinormal frequency and effective potential respectively.
Here $f(r)$ is determined by the metric, while the black hole's event horizon $r_h$ and cosmological horizon $r_c$ are often determined by $f(r_h)=0$ and $f(r_c)=0$ respectively.
We proceed by noting that the above form of the master equation can generally be applied to a wide variety of metrics. However, it is not applicable to the case of either rotating black holes or those in terms of Eddington–Finkelstein coordinates.
The specific forms of the master equation in those cases will be discussed as we handle the corresponding topics in the following sections.
Usually, the asymptotic behavior of the effective potential at the boundary is relatively simple, and therefore in the present work, we will just assume that the effective potential vanishes at the horizon and approaches a constant at infinity $V(\infty)\equiv V_\infty=\text{const}$.

In order to solve a differential equation, one also has to determine the associated boundary conditions.
In the case of quasinormal modes, the boundary conditions are determined by the characteristics of the black hole horizon and infinity.
As the apparent horizon of a black hole is a one-way membrane which prohibits the outgoing light rays to travel across it, the boundary conditions defining the quasinormal modes are that the solutions should be ingoing at the horizon and purely outgoing at infinity, namely
\bq
\lb{qnmbc}
\Phi\sim 
\left\{\begin{array}{ccc}
e^{i\bar{\omega} r_*}     &  r\rightarrow\infty &(\text{for asymptotically flat spacetime}) \cr\\
e^{i{\omega} r_*}&  r\rightarrow r_c & (\text{for asymptotically dS spacetime})  \cr\\
e^{-i\omega r_*}    &  r\rightarrow r_h & (\text{for both spacetimes})  
\end{array}\right. . 
\eq
where $\bar{\omega}$ is defined to be $\bar{\omega}\equiv \sqrt{\omega^2-V_\infty}$, so that $\bar{\omega}=\omega$ as $V_\infty=0$.
If the form of the effective potential is not simple enough, one usually has to resort to numerical methods.
Eqs.(\ref{qnmeq}) and (\ref{qnmbc}) furnish a eigenvalue problem, where its boundary conditions are defined at $r_*\rightarrow\pm\infty$ in terms of the tortoise coordinate.
We note that above discussions are for the asymptotically flat and dS spacetimes, while the boundary conditions for the asymptotically AdS spacetime and the angular part of the master equation will be addressed below.

Similar to other methods based on series expansion, we first change the domain of the radial variable to a finite range by introducing $z=z(r)$.
For convenience, we choose the new domain to be $z\in[0,1]$, and the boundaries correspond to the points $z=0$ and $z=1$ respectively.
However, since the wave function is also transformed accordingly, the boundary conditions Eq.(\ref{qnmbc}) will also be affected.
In this regard, we rewrite the wave function in the form $\Phi=A(z)R(z)$, where $A(z(r))$ carries the asymptotic behavior of the wave function at the boundary determined in Eq.(\ref{qnmbc}).
As a result, $R(z)$ is expected to be well behaved and in particular, regular at the boundary.
To be specific, the master equation reads
\bqn
\lb{3}
{\cal H}(\omega,z)R(z)=0 ~,
\eqn
with
\bqn
\lb{4}
R(z=0)=C_0~~~\text{and}~~~R(z=1)=C_1 ~.
\eqn
Here ${\cal H}(\omega,r)$ is a linear operator, determined by Eq.(\ref{qnmeq}), which depends on $\omega$ and $z$.
$C_0$ and $C_1$ are constants since the asymptotic part of the wave function has been factorized.
It is convenient to further introduce
\bqn
\lb{5}
F(z)=R(z){z(1-z)} ~,
\eqn
where $z(1-z)$ can be replaced by other smooth functions which only attains zero at $z=0$ and $z=1$.
We note that for the case of AdS spacetimes, since $R(r\rightarrow\infty)\rightarrow 0$, the transformation Eq.(\ref{5}) can be replaced by $F(z)=R(z)(1-z)$ where $z=r_h/r$.

Subsequently, that the master equation can be rewritten regarding $F(z)$
\bqn
\lb{qnmeqf}
{\cal G}(\omega,z)F(z)=0 ~,
\eqn
with the boundary conditions in a even simpler form
\bqn
\lb{qnmbcf}
F(z=0)=F(z=1)=0 ~.
\eqn
Here the linear operator ${\cal G}(\omega,z)$ is subsequently derived by the form of ${\cal H}(\omega,z)$ and Eq.(\ref{5}).
We note that, as a numerical trick, Eq.(\ref{5}) is not an absolutely necessary procedure.
In practice, the resultant wave function no longer possesses a boundary condition involving any undetermined constant.
Furthermore, when $C_0$ and $C_1$ turn out to be quite distinct in their values, the factor ${z(1-z)}$ will help to suppress the function values at the boundary points as well as in their vicinity.
Intuitively, the only scenario when the procedure does not work as intended is that the form of the wave function becomes more fluctuating due to the factor ${z(1-z)}$, for some particular reason, and the subsequent Taylor expansion becomes less efficient.
But since we do not have a good estimation of the form of the wave function at this point, the introduction of Eq.(\ref{5}) helps to maintain the resultant equation being more simplified, to the least.
The existing numerical results are mostly found to be consistent with the above speculations.

As for the next step, we descretize the master equation Eq.(\ref{qnmeqf}) by introducing $N$ interpolation points $z_i$ with $i=1, 2, \cdots, N$ for the interval $z\in[0,1]$.
Subsequently, the wave function $F(z)$ is also represented by descrete values $f_i=F(z_i)$ on the grids.
As a result, one discretizes the differential equation Eq.(\ref{qnmeqf}) and rewrites it in terms of $f_i$.
For the present algorithm, various derivatives of the wave function at $z_i$ are obtained by inverting a $N\times N$ dimensional matrix~\cite{agr-qnm-lq-matrix-01}, so that the information on the function values of the entire interval is utilized.
Besides, owing to the fact that the above inverting process is formal, in practice it is carried out before the real numerical calculations, and therefore it does not have any negative impact on the efficiency of the method.  
The resulting matrix equation can be formally written as
\bqn
\lb{7}
\bar{\cal M}{\cal F}=0 ~,
\eqn
where $\bar{\cal M}$ represents the descretized version of the operator ${\cal G}(\omega,z)$, it is a $N\times N$ matrix which still is a function of the quasinormal frequency $\omega$, and
\bqn
{\cal F}=(f_1,f_2,\cdots,f_i,\cdots,f_N)^T
\eqn
is a column vector composed of $f_i$.
The boundary condition Eq.(\ref{qnmbcf}) implies that
\bqn
\lb{8}
f_1=f_N=0~.
\eqn
We make use of the above condition to replace the first and last line of the matrix $\bar{\cal M}$.
The reason behind this choice is to select a line or column in the original equation which makes use of the least amount of information in terms of grid points.
Subsequently, Eq.(\ref{7}) becomes
\bqn
\lb{qnmeqMatrix}
{\cal M}{\cal F}=0~,
\eqn
where the element ${\cal M}_{k,i}$ of the matrix ${\cal M}$ is defined by
\bq
\lb{10}
{\cal M}_{k,i}= 
\left\{\begin{array}{cc}
\delta_{k,i},     &  k=1~\text{or}~N \cr\\
\bar{\cal M}_{k,i}, &  k=2,3,\cdots,N-1
\end{array}\right. ~.
\eq
The matrix equation indicates that ${\cal F}$ is the eigenvector of ${\cal M}$, which implies
\bqn
\lb{qnmDet}
\det\left({\cal M}(\omega)\right)\equiv\left|{\cal M}（(\omega)\right|=0~.
\eqn
Eq.(\ref{qnmDet}) is a non-linear algebraic equation of the quasinormal frequency $\omega$, which can be solved by any standard nonlinear equation solver.

In Refs.~\cite{agr-qnm-lq-matrix-02,agr-qnm-lq-matrix-03}, the above algorithm is employed to descretize the master equation, Eq.(\ref{qnmeqf}).
Alternatively, one may also employ other numerical approaches concerning numerical differentiation, such as differential quadrature method~\cite{math-algorithm-dq-01,math-algorithm-dq-02,math-algorithm-dq-03}, Runge-Kutta method, among others. 
Regarding the finite difference formulas, if Newton's difference quotient is employed in Eq.(\ref{qnmeqf}), the resulting matrix ${\cal M}(\omega)$ will be ``almost" diagonal since only the elements ${\cal M}_{k,i}$ with $i=k, k\pm 1$ are non-vanishing.
Subsequently, the resulting wave function can be obtained via the iterative method, similar to those of the continued fraction method and the HH method.
However, as mentioned above, a sparse matrix indicates that the corresponding discretizing scheme does not fully make use of the information on the entire interval of $z\in[0,1]$.
On the other hand, for several different methods, such as the continued fraction method and HH method, the series expansion of the wave function is only carried out on a single radial position, for instance, the horizon.
From a different viewpoint, in fact, the resulting algebraic equation can also be rephrased in the form of a matrix equation, Eq.(\ref{qnmeqMatrix}).
Owing to the fact that the iterative relation only involves a few expansion coefficients, the corresponding matrix is also rather sparse.
Therefore, for a given $N$, the matrix method is related to a Taylor expansion also of order $N$ which subsequently provides better precision. 
Moreover, one may even seek out an optimized grid distribution which provides a higher resolution for the region, where the variation of the wavefunction is more dramatic, by inserting more grid points. 
In this regard, the present algorithm can be viewed as an improvement over many existing approaches.

The following section is devoted to more detailed discussions of various applications.

\section{III. The application of the method in the calculations of quasinormal frequency}
\renewcommand{\theequation}{3.\arabic{equation}} \setcounter{equation}{0}

While the general strategy for the evaluation of the black hole quasinormal modes presented in the previous section is concise and straightforward, many particular aspects, such as those related to the boundary conditions and specific characteristics of the master equations, might bring up further complication for particular problems.
In the present section, we discuss in details several specific cases for the applications of the matrix method.
In particular, our discussions enumerate the boundary conditions for the black hole metrics in asymptotically flat, dS, and AdS spacetimes, as well as the associated ansatz for the method of separation of variables.
The background metrics concern static, rotational, and extreme black holes.
The algorithm presented in the following does not require the existence of an analytic form of the tortoise coordinate.
We also comment on the case of numerical black hole metric and those where the master equation is a system of coupled ordinary differential equations.

We first address the issue concerning different boundary conditions owing to different asymptotical properties of the spacetimes.
Usually, the spacetimes can be classified according to their asymptotical behavior at infinity, such as asymptotically flat, dS, and AdS spacetimes.
It is known that the effective potential $V(r)$ vanishes at infinity, in the case of asymptotically flat spacetime.
On the other hand, it diverges at infinity for the asymptotically dS and AdS spacetimes.
However, for the dS case, there exists a cosmological horizon $r_c$ between the observer and the infinity, which is also a Cauchy one-way membrane.
An observer staying between $r_h$ and $r_c$ will not receive any signal from inside of event horizon, neither from outside of the cosmological horizon. 
Therefore, the physically valid domain is $r_h\le r \le r_c$ for the dS spacetimes, while it is $r_h \le r \le \infty$ for for the asymptotically flat and AdS spacetimes.

For extreme black holes, the associated tortoise coordinate presents a different feature near the black hole horizon, from those of non-extreme black holes.
This is because, for an extreme black hole, one requires $f'(r_h)=0$, which implies the Taylor expansion of $f(r)$ around the horizon does not contain the linear term.
Subsequently, as shown below, the leading contribution in the tortoise coordinate becomes different.

For rotating black holes, an additional equation is implied whose physical content is associated with the angular momentum quantum number.
In this case, one usually encounters a system of two coupled master equations with eigenvalues $\omega$ and $\lambda$, corresponding to the radial and angular parts of the wave functions.
The latter restores to a simple case when the parameter of the black hole related to angular asymmetry vanishes.
In particular, if in this case, the black hole metric restores to a static spherical one, one finds $\lambda=\ell(\ell+1)$, namely, the angular momentum quantum number. Meanwhile, the angular part of the master equation becomes decoupled, and the angular wavefunction becomes spherical harmonics.
On the other hand, while the angular part of the master equation is coupled to the radial counterpart, its boundary condition is the usual periodic boundary condition.

In addition, one frequently encounters two difficult scenarios in realistic numerical studies.
The first one involves the difficulty in obtaining an analytic expression $r_*(r)$ for the tortoise coordinates, and subsequently, it becomes less straightforward write down the boundary condition in an analytic fashion.
The second case is apparently even more severe, which concerns the numerical black hole metric.
Fortunately, as we are about to argue below, the matrix method is quite versatile in handling these subtle situations.
In fact, it is relatively intuitive to understand the reason.
The critical procedure of the matrix method is to discretize continuous functions into discrete values on the grids.
Naturally, as a result, the implementation of the method does not rely on analytic expressions of the tortoise coordinate, as shown in the following subsection.
Moreover, the method neither relies on the analytic form of the black hole metric itself.
It is rather straightforward to slightly alter the procedure of the method to adopt numerical black hole metric.
This is not the case for other approaches such as the HH method, where the numerical difficulty is enhanced by the requirement of precious values for high order derivatives of the metric.
Last but not least, for the difficulty concerning numerical tortoise coordinate, one does not require an explicit form of the $r_*$ but its asymptotical behavior, and therefore, one can merely expand the form of $r_*$ near the boundary and extract the desired boundary conditions.

\subsection{A. The boundary conditions}

In the following, we write down explicit forms of the boundary condtions for five specific cases regarding different asymptotic spacetime and boundaries:

\textbf{Case 1.}  For non-exterme black holes, near the horizon, if $f(r)=\hat{f}_1(r-r_h)+\frac{\hat{f}_2}{2}(r-r_h)^2+{\cal O}(r-r_h)$ (where $\hat{f}_i\equiv f^{(i)}(r_h)$ and ${\cal O}$ represents terms of order higher than $(r-r_h)^2$), we have
\bqn
\lb{12}
r_*&=&\int\frac{dr}{f(r)}=\int\left(\frac{1}{\hat{f}_1(r-r_h)}-\frac{\hat{f}_2}{2\hat{f}_1^2}+{\cal O}\right)dr\nb\\
&=&\frac{\ln\left|r-r_h\right|}{\hat{f}_1}-\frac{\hat{f}_2}{2\hat{f}_1^2}(r-r_h)+{\cal O}~,
\eqn
and it implies that the boundary condition at the horizon is given by
\bqn
\lb{13}
\Phi\sim e^{-i\omega r_*}\sim B_a\equiv \left(r-r_h\right)^{-\frac{i\omega}{\hat{f}_1}}~.
\eqn

\textbf{Case 2.}  For exterme black holes, near the horizon, if $f(r)=\frac{\hat{f}_2}{2}(r-r_h)^2+\frac{\hat{f}_3}{6}(r-r_h)^3+\frac{\hat{f}_4}{24}(r-r_h)^4+{\cal O}(r-r_h)$, where ${\cal O}$ are terms of order higher than $(r-r_h)^4$, we find
\bqn
\lb{14}
r_*&=&\int\frac{dr}{f(r)}=\int\left(\frac{2}{\hat{f}_2(r-r_h)^2}-\frac{2\hat{f}_3}{3\hat{f}_2^2(r-r_h)}+\frac{4\hat{f}_3^2-3\hat{f}_2\hat{f}_4}{18\hat{f}_2^3}+{\cal O}\right)dr\nb\\
&=&-\frac{2}{\hat{f}_2(r-r_h)}-\frac{2\hat{f}_3\ln\left|r-r_h\right|}{3\hat{f}_2^2}+\frac{4\hat{f}_3^2-3\hat{f}_2\hat{f}_4}{18\hat{f}_2^3}(r-r_h)+{\cal O}~,
\eqn
so for extreme black holes, the boundary condition at horizon is given by
\bqn
\lb{15}
\Phi\sim e^{-i\omega r_*}\sim  B_b\equiv\left(r-r_h\right)^{\frac{2i\omega \hat{f}_3}{3\hat{f}_2^2}}e^{\frac{2i\omega}{\hat{f}_2(r-r_h)}}~.
\eqn

\textbf{Case 3.} In asymptotically flat spacetimes, for $r\to \infty$, if $f(r)=1+A r^\alpha+{\cal O}$, where $\alpha<0$ and ${\cal O}$ represents higher order terms less significant than $r^{\alpha}$, we have
\bqn
\lb{16}
r_*&=&\int\frac{dr}{f(r)}=\left.\int\left(1-Ar^{\alpha}+{\cal O}\right)dr\right|^{r\rightarrow \infty}=r-\frac{A}{1+\alpha}r^{1+\alpha}+{\cal O}~.
\eqn
Thus the boundary condition at infinity is given by
\bqn
\lb{17}
\Phi\sim e^{i\bar{\omega} r_*}\sim B_0\equiv e^{i\bar{\omega}\left(r-\frac{A}{1+\alpha}r^{1+\alpha}\right)} ~.
\eqn
and the second term could be significant when $\alpha>-1$.

\textbf{Case 4.} In asymptotically de Sitter spacetimes, according to Case 1 and Case 2, one obtains similar results by replacing $r_h$ with $r_c$.  
Therefore, corresponding Eq.(\ref{13}), near the non-extreme cosmological horizon, the boundary condition is
\bqn
\lb{18}
\Phi\sim e^{i\omega r_*}\sim B_1\equiv\left(r_c-r\right)^{\frac{i\omega}{\bar{f}_1}} ~.
\eqn
On the other hand, for the extreme cosmological horizon, we have
\bqn
\lb{19}
\Phi\sim e^{i\omega r_*}\sim  B_2\equiv\left(r_c-r\right)^{-\frac{2i\omega \bar{f}_3}{3\bar{f}_2^2}}e^{\frac{2i\omega}{\bar{f}_2(r_c-r)}} ~,
\eqn
where $\bar{f}_i\equiv f^{(i)}(r_c)$.

\subsection{B. Separation of variables}

By making use of the above-obtained asymptotic behavior, one can write down the appropriate forms for the wave functions by the method of separation of variables.

For non-extreme black holes in asymptotically flat spacetimes, we have
\bq
\lb{21}
\Phi(r)=e^{i\bar{\omega}\left(r-\frac{A}{1+\alpha}r^{1+\alpha}\right)}\left(r-r_h\right)^{-\frac{i\omega}{\hat{f}_1}}r^{\frac{i\omega}{\hat{f}_1}}R(r) ~,
\eq
where the term $r^{\frac{i\omega}{\hat{f}_1}}$ is introduced to cancel out the effect of the factor $\left(r-r_h\right)^{-\frac{i\omega}{\hat{f}_1}}$ at infinity while it gives a rather smooth contribution at the horizon.
Thus it can be readily verified that the resultant expression, Eq.(\ref{21}), indeed possesses the desired asymptotic forms at $r\to r_h$ and $r\to\infty$ found above in Eq.(\ref{13}) and (\ref{16}).
Likewise, for extreme black holes, we have
\bq
\lb{22}
\Phi(r)=e^{i\bar{\omega}\left(r-\frac{A}{1+\alpha}r^{1+\alpha}\right)}\left(r-r_h\right)^{\frac{2i\omega \hat{f}_3}{3\hat{f}_2^2}}e^{\frac{2i\omega}{\hat{f}_2(r-r_h)}}r^{-\frac{2i\omega \hat{f}_3}{3\hat{f}_2^2}}e^{-\frac{2i\omega}{\hat{f}_2r}}R(r)
\eq

As discussed above, for black holes in asymptotically de Sitter spacetime, at the event horizon it is required that $\hat{f}_1\not=0$ and $\hat{f}_1=0$ for the non-extreme and extreme cases respectively.
Similarly, at the cosmological horizon, we have $\bar{f}_1\not=0$ and $\bar{f}_1=0$ respectively for non-extreme and extreme scenarios. 
Therefore, the boundary conditions in de Sitter spacetimes are satisified by assuming the following form for the wave functions
\bq
\lb{23}
\Phi(r)= 
\left\{\begin{array}{cc}
B_aB_1R(r),     &  \text{when}~\hat{f}_1\not=0~\text{and}~\bar{f}_1\not=0 \cr\\
B_aB_2R(r),     &  \text{when}~\hat{f}_1\not=0~\text{and}~\bar{f}_1=0 \cr\\
B_bB_1R(r),     &  \text{when}~\hat{f}_1=0~\text{and}~\bar{f}_1\not=0 \cr\\
B_bB_2R(r),     &  \text{when}~\hat{f}_1=0~\text{and}~\bar{f}_1=0 
\end{array}\right. .
\eq

For asymptotically anti-de Sitter spacetime, the Horowitz-Hubeny method usually employs the ingoing Eddington-Finkelstein coordinates $(v,r)$ with $v=r_*+t$.
In what follows, we will also follow this convention.
The corresponding master equation reads
\bqn
\lb{24}
f(r)\frac{d^2\Phi(r)}{dr^2}+\left[\frac{df(r)}{dr}-2i\omega\right]\frac{d\Phi(r)}{dr}-U(r)\Phi(r)=0
\eqn
where $U(r)=V(r)/f(r)$, and since the potential is divergent at infinity, the boundary condition requires $\Phi$ being a constant at event horizon while vanishing at infinity.

For rotating black holes, one has to deal with the angular part of the master equation, which possesses the following form, similar to that of the radial part of the master equation:
\bqn
\lb{26}
(1-u^2)\frac{d}{du}\left[(1-u^2)\frac{dY(u)}{du}\right]-{\cal W}(u,\omega,\lambda)Y(u)=0 ~.
\eqn
The boundary conditions are at $u=\pm1$ where one defines $W^2_\pm ={\cal W}(u=\pm 1,\omega,\lambda)$.
The asymptotic solutions at the boundary are given by
\bq
\lb{27}
Y(u)\sim 
\left\{\begin{array}{cc}
\left|1-u\right|^{\pm\frac{W_+}{2}}    &  u\rightarrow  1 \cr\\
\left|1+u\right|^{\pm\frac{W_-}{2}}    &  u\rightarrow -1
\end{array}\right. .
\eq
As one requires that the angular part of the wave function $Y(u)$ always remains finity, it can be assumed to have the following form
\bq
\lb{28}
Y(u)=\left|1-u\right|^{\left|\frac{W_+}{2}\right|}\left|1+u\right|^{\left|\frac{W_-}{2}\right|}S(u) .
\eq
In the above expressions for the wave function by separation of variables through Eqs.(\ref{21})-(\ref{28}), we have implied that the functions $R(r)$ and $S(u)$ are finite and mostly well-hehaved for their entire domain regarding $r$ and $u$. 
That is to say, $r_h\le r < \infty$ for asymptotically flat and anti-de Sitter spacetimes, $r_h\le r\le r_c$ for asymptotically de Sitter spacetimes, and $-1\le u\le 1$ for angular part of the master equation.
This is because the asymptotic form of the wave function at the boundary has already be factorized and therefore effectively extracted from the l.h.s. of Eqs.(\ref{21})-(\ref{28}).

Last but not least, we carry out another coordinate transformation to conveniently convert the domain of $r$ and $u$ into $[0,1]$. 
For asymptotically flat and anti-de Sitter black hole spacetimes, the new coordinate could be chosen as $x=1-\frac{r_h}{r}$ or $z=r_h/r$, where $r_h$ is event horizon, so that $x=0$ and $z=1$ are at event horizon, while $x=1$ and $z=0$ represent infinity. 
For asymptotically de Sitter black hole spacetime, due to the existence of the cosmological horizon $r_c$, we choose $y=\frac{r-r_h}{r_c-r_h}$ so that $y=0$ represents the event horizon while $y=1$ indicates the cosmological horizon. 
For the angular coordinate, we introduce $\nu=\frac{1+u}{2}$ so that one has $\nu=0$ and $\nu=1$ for $u=-1$ and $u=1$ respectively. 

By following the above procedure, the transformed master equation, as well as its boundary conditions, are expressed in Eq.(\ref{3}) and Eq.(\ref{4}).
Subsequently, one may further proceed with the calculations of the quasinormal frequency concerning the matrix equation as discussed in Section II.

\subsection{C. Master equation concerning a system of coupled equations}

In the above discussions, we have assumed that the master equation can be expressed in terms of a system of decoupled equations, as we have treated radial and angular parts of the master equation separately.
In practice, such simplification might not always be possible.
As an example, in the study of quasinormal modes of a massive vector field, the resultant system of equations might be coupled.
However, a system of coupled ordinary differential equations does not particularly pose a difficulty for the matrix method.
This is because, in this case, one only need to write the system of coupled equations in terms of a matrix equation of larger size.
In order words, if a system of $n$ equations cannot be decoupled, the complication is formally demonstrated as one has to handle a  matrix equation of the size $nN\times nN$, where $N$ is the number of interpolation points introduced above.
As an example, in the case of $n=2$, we have
\bqn
\lb{29}
{\cal M}_a{\cal F}_a={\cal B}_b{\cal F}_b~,\nb\\
{\cal M}_b{\cal F}_b={\cal B}_a{\cal F}_a ~,
\eqn
where ${\cal B}$ is also a $N\times N$ matrix.
In this case, one may either rewrite the above equations in terms of a double-sized matrix equation, namely,
\bq
\lb{30}
{\cal M}_{n}{\cal F}_n\equiv\left(
  \begin{array}{cc}
    {\cal M}_a & -{\cal B}_b\\
    -{\cal B}_a & {\cal M}_b\\
  \end{array}
\right)
\left(
  \begin{array}{c}
    {\cal F}_a\\
    {\cal F}_b\\
  \end{array}
\right)=0 ~,
\eq
which can be readily solved as before.
The only disadvantage is that the matrix on the l.h.s. of Eq.(\ref{30}) might be a sparse matrix, which subsequently implies a loss of computational efficiency.

Alternatively, one can either rewrite Eq.(\ref{30}) as
\bqn
\lb{31}
\left({\cal M}_a{\cal B}_a^{-1}{\cal M}_b-{\cal B}_b\right){\cal F}_b=0 ~.
\eqn
or
\bqn
\lb{32}
\left({\cal M}_b{\cal B}_b^{-1}{\cal M}_a-{\cal B}_a\right){\cal F}_a=0 ~.
\eqn
Here the expressions evolve the evaluations of  ${\cal B}_a^{-1}$ or ${\cal B}_b^{-1}$, when the relevant inverse does exist.
To solve Eq.(\ref{31}) or Eq.(\ref{31}) is to find the roots of a nonlinear equation of $\omega$, which is associated with $N\times N$ matrices.
The second approach may potentially improve the efficiency.

\subsection{D. Code implementation}

The code implementation has been carried out in {\it Mathematica}.
It takes advantage of the efficiency of existing packages on matrix manipulation as well as nonlinear equation solver.
The source code for Schwartzchild black holes in asymptotically flat, (anti-)de Sitter (AdS/dS) spacetimes, as well as Kerr and Kerr-Sen black hole metrics,  have been released publicly~\cite{agr-qnm-lq-matrix-02,agr-qnm-lq-matrix-03}.
In this subsection, we discuss a few technical aspects concerning the code implemented of the algorithm regarding {\it Mathematica}.

Firstly, {\it Mathematica} sometimes is not very efficient in handling decimals, especially when decimals are further processed as arguments of internal functions, such as trigonometric, exponential functions.
Therefore, in regard to the optimization of the computational time, one should avoid coding in decimals.
As a example, one may utilize the following {\it Macro} to transform decimal output of a function ``TU" into fraction:
\bqn
\lb{33}
\text{YOULI}[\text{TU}\_] := \text{Rationalize}[\text{Chop}[\text{Expand}[\text{N}[\text{TU},\text{precs}]], 10^{-\text{precs}}], 10^{-\text{precs}}] ~,
\eqn
where ``precs" is the desired precision of the function ``TU". 
If one assigns $precs=100$, the function output will be expressed in terms of fraction truncated to ten significant figures.

Using {\it Mathematica}, there are usually two methods to find the roots of Eq.(\ref{qnmDet}), which is essentially an algebraic equation.
They are {\it FindRoot} and {\it NSolve}.
The former is to search for a root of the function, starting from the initial point provided by the user, while the latter attempts to enumerate all existing roots of the function. 
For the first case, one has to find an appropriate initial guess for the quasinormal frequency, which can be achieved by the following approaches.
(1) One can make use of the result of another less accurate method, such as P\"oshl-Theller potential approximation, as the input of the {\it FindRoot} command.
(2) When the black hole metric in question restores to a more straightforward case by continuously varying its parameters, where the quasinormal frequency is already known for the latter, one may obtain the desired results by gradually changing the parameters of the metric to the relevant values.
We note that the above approaches may not work as intended for the case where the quasinormal frequency possesses a bifurcation in the parameter space.
However, this problem exists in general for a variety of methods, where the eigenvalue problem is expressed in terms of nonlinear equations.
(3) Also, one may make use of the {\it NSolve} package to calculate the quasinormal frequencies.
In principle, one may find quasinormal frequencies related to different quantum numbers by a single root-finding process.
However, the challenge is how to pick out real physical roots from the numerical results.
One should proceed with care, and compare the numerical values with those obtained from cases of known black hole metrics.
Moreover, the following rule of thumb may provide some general guidance.
For some cases, in particular, those where the resultant algebraic equation, Eq.(\ref{qnmDet}), is merely an $N$-th order polynomial, one can exhaust all the roots via {\it NSolve} command.
Among these roots, some of them are not physical, and they appear because of certain symmetry of the equation associated with a particular choice of interpolation points.
These unphysical roots can be observed as one varies the number of interpolation points, as the roots corresponding to physical quasinormal frequency will always persist and approaches its limit as $N\to \infty$.
Nonetheless, to properly remove all the unphysical roots is not a simple task.
The situation becomes even more complicated when Eq.(\ref{qnmDet}) is highly nonlinear.
In this regard, the above mentioned {\it Macro} can be employed to alleviate the situation.

Apart from the above two methods, the matrix method can also be adapted to an iterative scheme.
This third method is meaningful especially for situations where neither {\it FindRoot} nor {\it NSolve} is shown to be efficient.
In general, the iterative relation can be written as follows.
\bqn
\lb{34}
\omega_{k+1}={\cal Y}\left(\omega_k,\left|{\cal M}(\omega)\right|\right) ~,
\eqn
where the quasinormal frequency of the $(k+1)$-th interation, $\omega_{k+1}$, is determined by the the $k$-th result.
The operator ${\cal Y}$ is a functional of the determinant satisfying $\omega={\cal Y}\left(\omega,\left|{\cal M}\right|\right)$ for $\omega$ being the root of Eq.(\ref{qnmDet}).
By appropriate choosing the specific form of ${\cal Y}$, one obtains the result of quasinormal frequency by repeatedly using Eq.(\ref{34}).
The primary challenge of this approach to find a proper iteration relation regarding ${\cal Y}(x,y)$ which efficiently leads to a convergent result.

In our code implementation in Refs.\cite{agr-qnm-lq-matrix-02,agr-qnm-lq-matrix-03}, we have evenly divided the domain $[0,1]$ of $z$.
In practice, it might be more favorable to utilize nonuniform distribution which introduces more interpolation points in the region where the effective potential changes more radically.
As a matter of fact, this is a similar issue that one encounters in other numerical schemes, such as the differential quadrature method and the Runge-Kutta method.

\subsection{E. Additional considerations}

There are a few more issues or topics which have not been explored in the existing studies but worth mentioning.

Apart from the asymptotically AdS spacetime, our analysis has been mostly carried out in radial coordinate $r$.
However, sometimes it can be more convenient to phrase the formalism in ingoing Eddington–Finkelstein coordinates.
In this case, the corresponding boundary condition also has to be modified accordingly
\bqn
\lb{34}
\Phi\sim 
\left\{\begin{array}{cc}
e^{i\left(\omega+\tilde{\omega}\right) r_*}     &  r\rightarrow\infty ~\text{or}~r\rightarrow r_c\cr\\
\Phi_0    &  r\rightarrow r_h
\end{array}\right. ~.
\eqn
where $\tilde{\omega}=\sqrt{\omega^2-V_\infty}$ for $r\rightarrow\infty$ in asymptotically flat spacetimes and $\tilde{\omega}=\omega$ for $r\rightarrow r_c$ in asymptotically dS spacetimes.
We note that the master equation can also be derived in outgoing Eddington-Finkelstein coordinates in a similar fashion.

In practice, instead of carrying out the transformation Eq.(\ref{5}) and rewrite the boundary condition as Eq.(\ref{qnmbcf}), the original boundary condition Eq.(\ref{4}) might be directly employed in the calculation.
As mentioned before, the primary motivation of introducing Eq.(\ref{5}) is when $C_0$ and $C_1$ are very distinct in their magnitudes.
It is understood that such a procedure is useful regarding the radial coordinate $r$ discussed in the present work.

In reality, the black hole is not a static object.
The topic on quasinormal modes for dynamic black holes is not only intriguing but also significant in practice.
For a dynamical black hole, event horizon, apparent horizon, infinite redshift surface are distinct concepts, and therefore the boundary condition of the quasinormal modes should be tackled with special care.

In certain modified gravity theories, the Lorentz invariance is broken.
Owing to tachyon, the concepts of the event horizon and apparent horizon have to be modified in these theories.
Recent studies have proven the existence of the universal horizon~\cite{agr-modified-gravity-horava-lw-06,agr-modified-gravity-horava-lw-08}, which is defined to be the boundary that even superluminal particles cannot escape to the future infinity.
As a part of the efforts to search for evidence of breaking Lorentz invariance, it is, therefore, interesting to investigate the quasinormal modes regarding universal horizon in the framework of the matrix method.

The notion of quasinormal is not restricted to black holes.
Quasinormal modes of compact stars have long aroused much attention of physicists.
The primary difference between the two phenomena, from a mathematical viewpoint, is the existence of the outgoing wave at the boundary of the star surface.
Therefore, it implies a more subtle condition to properly connect the wave functions from inside and outside the boundary.
Nonetheless, the matrix method can also be adapted to handle such problems appropriately.

Last but not least, we note that the quasinormal mode, by itself, is the eigenvalue of the secular equation.
The matrix method is merely the discretized version of the eigenvalue problem in question.
One of the alternatives for the eigenvalue problem is the shooting method, which solves a boundary value problem by reducing it to the that of an initial value problem.
In this context, one might also implement a specific version of the shooting method concerning the column vector, which is nothing but the discretized eigenfunction.

\section{IV. Concluding remarks}
\renewcommand{\theequation}{4.\arabic{equation}} \setcounter{equation}{0}

To summarize, in the present paper, a comprehensive survey on the applications of the matrix method for black hole quasinormal modes is presented.
It is shown that the proposed algorithm is independent of any particular form of analytic expression for the tortoise coordinate.
In fact, the method can even be employed to handle numerical black hole metric.
Moreover, it is argued that the present approach can readily be applied to the cases where the master equation involves a system of coupled equations. 
Our discussions cover various types of black hole metrics, master equations, and the corresponding boundary conditions.
The proposed method is reasonably accurate, efficient, as well as flexible for distinctive physical scenarios.
We look forward to carrying out further studies regarding several open questions addressed in the present survey.

\section*{\bf Acknowledgements}

We gratefully acknowledge the financial support from National Natural Science Foundation of China (NNSFC) under contract No.11805166,
as well as Brazilian funding agencies
Funda\c{c}\~ao de Amparo \`a Pesquisa do Estado de S\~ao Paulo (FAPESP),
Conselho Nacional de Desenvolvimento Cient\'{\i}fico e Tecnol\'ogico (CNPq),
and Coordena\c{c}\~ao de Aperfei\c{c}oamento de Pessoal de N\'ivel Superior (CAPES).

\bibliographystyle{h-physrev}
\bibliography{references_qian,references_etc}

\end{document}